\UseRawInputEncoding
\documentclass[
superscriptaddress,
bibnotes,
 amsmath,amssymb,
 aps,
pra,
twocolumn
]{revtex4-1}
\usepackage{graphicx}
\usepackage{dcolumn}
\usepackage{xr}
\usepackage{bm} 
\usepackage{float}
\usepackage{hyperref} 
\hypersetup{
	colorlinks=true,
	linkcolor=blue,
	citecolor=blue,
	filecolor=magenta,
	urlcolor=blue         
}
\usepackage{color,soul}
\usepackage{lineno} 

\renewcommand{\selectlanguage}[1]{}

\begin{document}

\title{Phonon decoherence produced by two-level tunneling states}

\author{Ryan O. Behunin}
\email[]{ryan.behunin@nau.edu}
\affiliation{Department of Applied Physics and Materials Science, Northern Arizona University, Flagstaff, AZ 86011, USA}
\affiliation{Center for Materials Interfaces in Research and Applications (<MIRA!), Flagstaff, AZ, USA}

\author{Taylor Ray}
\affiliation{Department of Applied Physics and Materials Science, Northern Arizona University, Flagstaff, AZ 86011, USA}
\affiliation{Center for Materials Interfaces in Research and Applications (<MIRA!), Flagstaff, AZ, USA}

\author{Dylan Chapman}
\affiliation{Department of Applied Physics and Materials Science, Northern Arizona University, Flagstaff, AZ 86011, USA}
\affiliation{Center for Materials Interfaces in Research and Applications (<MIRA!), Flagstaff, AZ, USA}

\author{Andrew J. Shepherd}
\affiliation{Department of Applied Physics and Materials Science, Northern Arizona University, Flagstaff, AZ 86011, USA}
\affiliation{Center for Materials Interfaces in Research and Applications (<MIRA!), Flagstaff, AZ, USA}

\author{Yizhi Luo}
\affiliation{Department of Applied Physics and E.L. Ginzton Laboratory, Stanford University, Stanford, CA 94305, USA}

\author{Peter T. Rakich}
\affiliation{Department of Applied Physics, Yale University, New Haven, CT 06520, USA}

\date{\today}

\begin{abstract}
Phonon modes within pristine crystalline resonators now routinely reach the quantum ground state. Such systems are attractive for quantum information science applications, as advanced fabrication and processing can enable relatively long quantum coherence times, and precision control can be realized through optical, electrical, or qubit coupling. In many state-of-the-art systems, the phonon lifetime is limited by disorder. In particular, native oxides or damaged `dead layers' at surfaces can host two-level tunneling states that lead to a particularly problematic form of dissipation that increases at lower temperatures. As mechanical losses are driven down in systems such as micro-fabricated bulk acoustic wave resonators, tunneling states are expected to emerge as the dominant mechanism for phonon decoherence. A quantitative description of these mesoscopic systems therefore requires a framework that captures interactions between a selected phonon mode and a large ensemble of TLS. Here, we derive a quantum master equation for this coupled system, permitting the phonon decoherence produced by two-level tunneling states to be calculated. As an example, we estimate the lifetime of a variety of quantum states within quartz micro-resonators hosting a thin surface layer of tunneling states. We find that the phonon coherence time is maximized at low temperatures, in spite of increased mechanical dissipation, and that phonon-TLS coupling can be reduced for modes with strain nodes at the surfaces. 
\end{abstract}

\maketitle

\section{Introduction}
On account of their long lifetime at cryogenic temperatures and interaction with optical and electrical signals, the precision control and measurement of phonons within crystalline media has enabled new ways to store, process, and transmit quantum information \cite{satzinger2018quantum,chu2017quantum,chu2018creation,sletten2019resolving,wollack2022quantum,yang2024mechanical,von2024engineering}.
This potential has driven extraordinary progress in the design, fabrication, and operation of ultra-high quality-factor mechanical resonators. Examples include the development of electrically coupled bulk quartz acoustic wave resonators reaching limits set by Rayleigh scattering \cite{goryachev2013}, optomechanical crystals supporting 5 GHz mechanical modes with 1.5 s lifetimes \cite{maccabe2020nano},  lithographically defined microresonators with surface-limited losses and $3\times 10^{18}$ fQ-products \cite{luo2025lifetime}, and operation of massive (as high as 494$\mu$g) optomechanical systems in the quantum ground state \cite{doeleman2023brillouin,diamandi2025optomechanical}, to name a few. 
Because fundamental phonon loss mechanisms within crystalline media sharply decrease with temperature \cite{maris1965absorption,maris1971interaction}, the performance of many of these emerging systems is ultimately limited by residual disorder, often found at surfaces \cite{goryachev2013,maccabe2020nano,luo2025lifetime,gruenke2024surface,gruenke2025surface}. In silicon, lithium niobate, and quartz systems, one particularly vexing form of dissipation and decoherence is produced by so-called two-level tunneling states \cite{phillips1987two,gruenke2024surface,gruenke2025surface,wollack2021loss,maccabe2020nano,cleland2024studying}. 

Two-level tunneling states (TLSs) are inherent to amorphous materials and are responsible for excess decoherence in a range of quantum systems \cite{behunin2016dimensional}. They are believed to arise from atoms, or groups atoms, that inhabit asymmetric double-well potentials of the type shown in Fig. \ref{fig:TLS-diagram}a. For crystalline systems they can be found on surfaces, such as within native oxides, or in `dead layers' where the lattice contains dislocations, damage, or impurities \cite{gruenke2024surface,gruenke2025surface}. And while the losses produced by TLSs can be `saturated' under intense excitation \cite{golding1973nonlinear,phillips1987two,martinis2005decoherence,behunin2017engineering,andersson2021acoustic} (or through phonon backreaction \cite{cleland2024studying}), in the weak excitation limit that is compatible with the operation of many quantum devices, these losses become more acute as the temperature is lowered. As advances in fabrication and processing continue to extend phonon coherence times, the consideration of TLSs in the quantum dynamics of phonons across a broad range of systems will become increasingly important. 

In this paper, we utilize the standard tunneling state model to derive the quantum master equation for a selected phonon mode interacting with an ensemble of TLSs. Our results complement prior work that has numerically and experimentally investigated the impact of a small collection of TLSs on the quantum dynamics of nanomechanical systems \cite{remus2009damping,remus2012damping,maccabe2020nano,cleland2024studying}. In contrast, we consider phonon modes that interact with a `large' ensemble of TLSs, such as bulk acoustic resonators, where the backreaction of the phonon on the TLS ensemble can be neglected. Under these conditions and taking the Markov approximation, we obtain a Lindblad-form phonon master equation with transition rates determined by the TLS-phonon losses. This result is amenable to an analytical solution that can be used to evaluate the evolution of arbitrary initial phonon states. As one example, we consider phonon Fock state superpositions, deriving expressions for the fidelity of state storage versus time, and applying these results to quartz $\mu$BAR systems as a function of temperature \cite{luo2025lifetime}. These results show that while resonant absorption by TLSs is suppressed with increased temperatures, i.e., saturated, decoherence is still rapid in this `high' temperature regime. This is largely due to the relative dominance of ``relaxation absorption'' in the phonon losses which cannot be saturated \cite{phillips1987two}. We also find that the phonon quantum state lifetime reaches a maximum at low temperatures, despite increases in the mechanical dissipation, and that coupling to TLSs within surface layers can be suppressed when the surface is a strain node. 

The paper is organized as follows: In Sec. \ref{sec: TLS model} we describe the TLS model, and discuss how it can be applied to crystalline systems. In Sec. \ref{sec: master equation} we derive the master equation for a selected phonon mode that interacts with an ensemble of TLSs. In Sec. \ref{sec: quantum dynamics} we show the exact solution to the phonon master equation, solve for the fidelity of Fock superposition state storage versus time, and estimate the  lifetime of a collection of quantum states proposed for qubit encoding in a quartz microresonator.  

\section{Two-level tunneling state model}
\label{sec: TLS model}

\begin{figure}
    \centering
    \includegraphics[width=\linewidth]{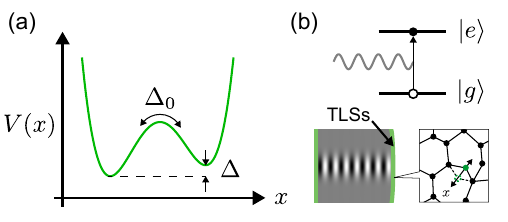}
    \caption{(a) double-well potential associated with two-level tunneling states, (b) TLSs, found on the surfaces of crystalline systems, modeled as two discrete energy levels that resonantly interact with phonons. }
    \label{fig:TLS-diagram}
\end{figure}

The coupled dynamics of a selected phonon mode with an ensemble of TLSs can be described by the standard two-level tunneling state model. In this model, TLSs are assumed to arise from atoms, or groups of atoms, that reside within asymmetric double-well configurational potentials \cite{phillips1987two} (Fig. \ref{fig:TLS-diagram}). While the microscopic origin and nature of TLS couplings is still not fully understood \cite{yu1988low}, this model makes quantitative predictions that are consistent with observations. 

The Hamiltonian for this model is given by 
\begin{align}
    H  = H_0 + H_{int} + H_{env}
\end{align}
where $H_0$ is the Hamiltonian describing the free-evolution of the TLS ensemble and the phonon,
\begin{align}
H_0 = \sum_j \frac{1}{2} E_j \sigma_{z,j} + \hbar \Omega b^\dag b.
\end{align}
Here, $\sigma_{z,j}$ is the z-component Pauli operator for the $j$th TLS, $E_j$ is  TLS energy,
and $H_{env}$ accounts for `environmental' degrees of freedom and their coupling to the TLS ensemble that lead to damping and thermalization. This environment is comprised of the remaining phonon modes in the system. The TLS-phonon interaction $H_{int}$ is given by 
\begin{align}
  H_{int} = \sum_j \hbar (g_j \sigma_{z,j} + g_{0,j}\sigma_{x,j})(b+b^\dag),  
\end{align}
where $\sigma_{x,j}$ is the x-component Pauli operator of the $j$th TLS, and $b$ and $b^\dag$ are the respective annihilation and creation operators for the phonon mode. The coupling rates $g_j$ and $g_{0,j}$ quantify the interaction between the selected phonon mode and the $j$th TLS. Expressed in terms of the level splitting $\Delta$, tunneling strength $\Delta_0$ (Fig. \ref{fig:TLS-diagram}), TLS energy $E= \sqrt{\Delta^2+\Delta_0^2}$, and the orthonormal functions for the phonon mode, these coupling rates are given by
\begin{align}
    g_j & \equiv \sqrt{\frac{1}{2 \hbar \Omega}}\frac{\Delta}{E} \boldsymbol{\gamma} : \boldsymbol{\xi}_{q}({\bf x}_j) 
    \\
     g_{0,j} & \equiv \sqrt{\frac{1}{2 \hbar \Omega}}\frac{\Delta_0}{E} \boldsymbol{\gamma} : \boldsymbol{\xi}_{q}({\bf x}_j)
\end{align}
where $\boldsymbol{\gamma}$ is the deformation potential tensor, comprised of longitudinal $\gamma_L$ and transverse components $\gamma_T$ (see Ref. \cite{anghel2007interaction} for further details). The tensor $\boldsymbol{\xi}_q$ is the strain associated with the normal mode amplitude ${\bf u}_q$ that satisfies the Christoffel equation and the orthogonality condition $\int d^3x \ \rho_0 {\bf u}^*_q \cdot {\bf u}_{q'} = \delta_{q q'}$, with $\rho_0$ denoting the mass density \cite{behunin2016dimensional}. Note that the phonon couplings to the $j$th TLS depend upon the value of the strain field at the TLS location ${\bf x}_j$. Therefore, the TLS coupling to a given phonon mode vanishes for TLS's located within strain nodes. For example, in systems with free (i.e., force-free) surfaces that host TLSs in a layer that is much thinner than the acoustic wavelength, TLS-phonon coupling is highly suppressed.   

For systems possessing a large ensemble of TLSs, the sum $\sum_j$ can be well approximated by the integral
\begin{align}
    \sum_j (\hdots) = \int d^3 x \int d\Delta d \Delta_0 \frac{P({\bf x})}{\Delta_0}(\hdots).
\end{align}
Here, the density of states $P({\bf x})$ is assumed to be constant within regions containing TLSs, slightly generalizing the tunneling state model to heterogeneous systems.

\section{Derivation of the phonon master equation}
\label{sec: master equation}

The quantum master equation provides a powerful method to calculate the dynamics of complex quantum systems in terms of time-dependent reduced density matrices \cite{breuer2002theory}. Here, we utilize the techniques of open quantum systems to derive a master equation for the phonon that accounts for interactions with an ensemble of thermalized TLSs to second order in the couplings $g_j$ and $g_{0,j}$.

We begin this derivation with the von Neumann equation in a rotating (i.e., interaction picture-like) frame defined by $\rho(t) = \exp\{ i(H_0'+H_{env})t/\hbar\} \rho_{Sch}(t) \exp\{-i(H_0'+H_{env})t/\hbar\}$ 
\begin{align}
    \dot{\rho}(t) = -\frac{i}{\hbar}[H_{int}(t),\rho(t)]
\end{align}
where $\rho_{Sch}$ is the Schrodinger picture representation of the density matrix. Here, $H_0' = H_0 + \hbar \delta\Omega b^{\dag} b$ where $\delta\Omega$ is a constant offset frequency that is used to cancel any frequency shifts produced by the TLSs (in the following $\delta\Omega =0$, see App. \ref{App: imaginary terms}). 
First, we obtain a formal solution for $\rho$ by direct integration
\begin{align}
    \rho(t) = \rho(t_i) -\frac{i}{\hbar} \int_{t_i}^t d\tau \ [H_{int}(\tau),\rho(\tau)].
\end{align}
Inserting this solution back into the von Neumann equation and tracing over all TLS and environment degrees of freedom, i.e., reservoir degrees of freedom (tr$_R$), we obtain the equation for the reduced density matrix $\rho_{S} = {\rm tr}_R\{\rho\}$ ($S$ for system)
\begin{align}
    \dot{\rho}_S(t) = -\frac{i}{\hbar} & {\rm tr}_R\{[H_{int}(t),  \rho(t_i)]\} 
   \\
   & -\frac{1}{\hbar^2} \int_{t_i}^t d\tau \ {\rm tr}_R\{ [ H_{int}(t),[H_{int}(\tau),\rho(\tau)]]\}. \nonumber
\end{align}
We drop the term containing $\rho(t_i)$, which we assume vanishes when averaged over TLS couplings and when the TLS are in thermal equilibrium \footnote{The impact of such a term may be important when the system contains a small number of TLSs and the averaged effect does not vanish.}. Owing to the relatively large size and short correlation time of the TLS ensemble, we assume: (1) the validity of the Born approximation, i.e, $\rho(t) = \rho_S(t) \otimes \rho_R$ where $\rho_R$ is the time-independent density matrix for the reservoir in thermal equilibrium, and (2) that the reservoir memory time is much shorter than the characteristic timescale for change in $\rho_S$ (i.e., the Markov approximation is valid). Under the Born-Markov approximation, the quantum master equation becomes
\begin{align}
    \dot{\rho}_S(t) = -\frac{1}{\hbar^2} \int_{t_i}^t d\tau \ {\rm tr}_R\{ [ H_{int}(t),[H_{int}(\tau),\rho_S(t)\otimes \rho_R]]\}. 
\end{align}

By defining $H_{int}(t) = \hbar O(t)x(t)$, where $x(t) = b \exp\{-i\Omega t\} + b^\dag \exp\{i\Omega t\}$ and $O(t) = \sum_j (g_j \sigma_{z,j}(t)+g_{0,j} \sigma_{x,j}(t))$, the coefficients of the master equation can be defined in terms of reservoir correlation functions, $ \mathcal{G}(t-\tau)$, as follows

\begin{align}
\label{Eq: ME-BM}
    \dot{\rho}_S(t) =
    - \int_{t_i}^t d\tau
    \big[ &
    \mathcal{G}(t-\tau)  x(t)x(\tau)\rho_S(t)
     \nonumber \\ 
      - &
    \mathcal{G}(\tau-t)  x(t)\rho_S(t) x(\tau)
   \nonumber \\ 
    - &
    \mathcal{G}(t-\tau)  x(\tau)\rho_S(t) x(t)
     \nonumber \\ 
     + &
    \mathcal{G}(\tau-t) \rho_S(t) x(\tau) x(t)
    \big]
\end{align}
where $\mathcal{G}(t-\tau) \equiv {\rm tr}_R\{\rho_R O(t) O(\tau) \}$, and we have used the cyclic property of the trace and the time-translation invariance of the assumed equilibrium state of the reservoir.

Equation \eqref{Eq: ME-BM} can be further simplified by using the definition of $x(t)$, taking $t_i \to -\infty$, and neglecting rapidly oscillating terms. After a change of variables, $t-\tau \to t'$, the integral terms contained in Eq. \eqref{Eq: ME-BM} take the form
\begin{align}
    \int_{0}^\infty dt' \ & \mathcal{G}(\pm t') A x(t) B x(t-t') C  = 
    \\
     &  \int_{0}^\infty dt' \mathcal{G}(\pm t')  
       \bigg( A b B b^\dag C e^{-i \Omega t'}
    + 
     A b^\dag B b C e^{i \Omega t'}
       \nonumber \\       
     & \quad \quad + A b B b C e^{-i \Omega (2 t- t')}
      + 
     A b^\dag B b^\dag C e^{i \Omega (2 t-t')}
     \bigg) \nonumber 
\end{align}
where $A$, $B$, and $C$ are assumed to be generic operators. Assuming that $\Omega$ is much greater than the TLS-induced decay rates, the last two, rapidly oscillating terms are suppressed and can be neglected according to the post-trace rotating wave approximation. 

Defining the susceptibility function  
\begin{align}
\label{Eq: chi}
    \chi_\pm(\Omega) = \int_0^\infty dt \ \mathcal{G}(\pm t) e^{i\Omega t}
\end{align}
and noting $\mathcal{G}(\pm t) = \mathcal{G}^*(\mp t)$ and $\chi_\pm(\Omega) = \chi^*_\mp(-\Omega)$,
 Eq. \eqref{Eq: ME-BM} can be expressed as
\begin{align}
       \dot{\rho}_S =
 2 {\rm Re}[\chi_+(\Omega)] b \rho_S b^\dag - 
 \chi_+(\Omega) b^\dag b\rho_S - \chi^*_+(\Omega)\rho_S b^\dag b \nonumber \\
 +2 {\rm Re}[\chi_+(-\Omega)] b^\dag \rho_S b - 
 \chi_+(-\Omega) b b^\dag \rho_S - \chi^*_+(-\Omega)\rho_S b b^\dag.
\end{align}
Note that imaginary parts of $\chi_{+}(\pm \Omega)$ lead to a perturbation of the phonon resonance frequency (see Appendix \ref{App: imaginary terms}). We assume that our rotating frame has been precisely adjusted to compensate for this frequency shift (as described in Sec. \ref{sec: master equation}), and from hereon neglect these terms. This last simplification brings us to the final form of the phonon master equation
\begin{align}
\label{Eq: ME-Lindblad}
       \dot{\rho}_S = &
\gamma_{\downarrow} \bigg( b \rho_S b^\dag -  \frac{1}{2} b^\dag b\rho_S - \frac{1}{2} \rho_S  b^\dag b \bigg)
\nonumber \\
& \quad +
\gamma_{\uparrow} \bigg( b^\dag \rho_S b -  \frac{1}{2} b b^\dag \rho_S - \frac{1}{2} \rho_S b b^\dag \bigg)
\end{align}
where the transition rates are 
\begin{align}
\label{Eq: down-rate}
    \gamma_\downarrow & = 2 {\rm Re}[\chi_+(\Omega)]
    \\
    \label{Eq: up-rate}
    \gamma_\uparrow & = 2 {\rm Re}[\chi_+(-\Omega)].
\end{align}
With the explicit evaluation of $\chi_+(\pm\Omega)$, the quantum dynamics of the phonon can be calculated. 

\subsection{Evaluation of the transition rates}

The transition rates $\gamma_{\uparrow}$ and $\gamma_{\downarrow}$ can be derived by evaluating the reservoir correlation function $\mathcal{G}(t)$ and taking the Fourier transform. Utilizing the definition of $O(t)$, the correlation function is given by 
\begin{align}
\label{Eq: correlation function}
\mathcal{G}(t) = \sum_{j} \bigg[ &
|g_j|^2 {\rm tr}_R\{ \rho_R \sigma_{z,j}(t) \sigma_{z,j}(0)\}
\nonumber \\
& +
|g_{0j}|^2 {\rm tr}_R\{ \rho_R \sigma_{x,j}(t) \sigma_{x,j}(0)\}
\bigg]
\end{align}
where the cross terms between $\sigma_{z,j}$ and $\sigma_{x,j}$ vanish in equilibrium and we assume ${\rm tr}_R\{ \rho_R O(t)\} = 0$. Using the adjoint Lindblad equation for the TLS ensemble, the correlation function $\mathcal{G}(t)$ can be evaluated and expressed in terms of equilibrium properties and relaxation timescales $T_1$ and $T_2$ (see Appendix \ref{App: TLS correlation functions})\cite{breuer2002theory}. This analysis provides
\begin{align}
\label{Eq: two-time correlation functions}
   {\rm tr}_R\{ \rho_R \sigma_{z,j}(t) \sigma_{z,j}(0)\} = & (1-w_j^2)e^{-\frac{|t|}{T_{1,j}}} + w_j^2
    \\
    {\rm tr}_R\{ \rho_R \sigma_{x,j}(t) \sigma_{x,j}(0)\} = &
    \frac{1}{2}\bigg[ (1+w_j)e^{iE_j t/\hbar} 
    \\
    \label{Eq: two-time correlation functions-2}
    & \quad \quad  + (1-w_j)e^{-iE_j t/\hbar}
    \bigg] e^{-\frac{|t|}{T_{2,j}}} \nonumber 
\end{align}
where $w_j$ is the thermal equilibrium value of $\sigma_{z,j}$
\begin{align}
    w_j = - \tanh\left( \frac{E_j}{2 k_B T}\right). 
\end{align}
These correlations functions satisfy the Pauli algebra (reproducing $1$ at $t=0$), yield the proper long-time thermal expectation values ($w_j^2$ and $0$, respectively), and capture the relaxation dynamics in the Markov approximation.

Combining Eqs. \eqref{Eq: correlation function}-\eqref{Eq: two-time correlation functions-2} \& \eqref{Eq: chi}, we find 
\begin{align}
    2 {\rm Re}[\chi_+(\pm \Omega)] = & \sum_j |g_j|^2 \bigg[\frac{2 T_{1,j}(1-w_j^2)}{1+\Omega^2 T_{1,j}^2}+2\pi \delta(\Omega) w_j^2 \bigg]
      \nonumber \\
    & + \sum_j  |g_{0j}|^2 \bigg[
     \frac{T_{2,j}(1+w_j)}{1+(\pm\Omega+E_j/\hbar)^2T_{2,j}^2}
    \nonumber \\
    &\quad \quad \quad \quad +
    \frac{T_{2,j}(1-w_j)}{1+(\pm\Omega-E_j/\hbar)^2T_{2,j}^2}
    \bigg].
\end{align}

The susceptibility above can be expressed in terms phonon damping rate produced by TLSs. Using Appendix E and F of \cite{behunin2016dimensional}, we have the following expressions for the dissipation rates produced by relaxation $\Gamma_{rel}$ and resonant processes $\Gamma_{res}$ 
\begin{align}
    \Gamma_{rel} & = - \sum_j  \hbar |g_j|^2 \frac{4 T_{1,j}\Omega}{1+\Omega^2 T_{1,j}^2} \frac{\partial w_j}{\partial E_j} 
    \\
    \Gamma_{res} & = - \sum_j |g_{0,j}|^2 \frac{2 T_{2,j}}{1+(\Omega-E_j/\hbar)^2 T_{2,j}^2} w_j.
\end{align}
Noting that $\beta(1-w_j^2)/2 = -\partial w_j/\partial E_j$, and assuming that $1/T_{2,j} \ll \Omega$, so that the $2 T_{2,j}/(1+(\Omega-E_j/\hbar)^2 T_{2,j}^2) \approx 2\pi \delta(\Omega-E_j/\hbar)$, Eqs. \eqref{Eq: down-rate} and \eqref{Eq: up-rate} become

\begin{align}
 \gamma_\downarrow  &= \Gamma_{rel} \frac{k_B T}{\hbar \Omega} + \Gamma_{res} (n_{th} +1)\\
  \gamma_\uparrow & = \Gamma_{rel} \frac{k_B T}{\hbar \Omega} + \Gamma_{res} n_{th} 
\end{align}
where $n_{th} = (\exp\{\hbar \Omega/k_B T\}-1)^{-1}$ is the thermal occupation number for the phonons. 
Note that these transition rates do not precisely satisfy detailed balance. However, for a broad range of systems, relaxation absorption only becomes significant to the transition rate when $k_B T \gg \hbar \Omega$. In this high temperature limit $n_{th} \approx n_{th}+1 \approx k_B T/\hbar \Omega$ yielding the approximate transition rates 
\begin{align}
\label{Eq: decay down}
 & \gamma_\downarrow \approx \Gamma (n_{th} +1)\\
 \label{Eq: decay up}
 & \gamma_\uparrow \approx \Gamma n_{th} 
\end{align}
where $\Gamma = \Gamma_{rel}  + \Gamma_{res}$. (see Ref. \cite{behunin2016dimensional} for derivations of $\Gamma$ in a a variety of system geometries). 
Using Eqs. \eqref{Eq: decay down} \& \eqref{Eq: decay up}, the master equation Eq. \eqref{Eq: ME-Lindblad} takes the standard Lindblad form obeying detailed balance. 

\section{Quantum dynamics of phonons}
\label{sec: quantum dynamics}

The quantum dynamics of any initial phonon state can be calculated using the known analytical solution to Eq. \eqref{Eq: ME-Lindblad} \cite{fujii2012quantum}. Using Lie-algebraic techniques and the disentangling theorem the closed-form solution to Eq. \eqref{Eq: ME-Lindblad} (with decay rates given by Eqs. \eqref{Eq: decay down} \& \eqref{Eq: decay up}) can be shown to be given by  
\begin{align}
\label{Eq: rho_red}
    \rho_S(t) = \frac{e^{\frac{\Gamma t}{2}}}{F} \sum_{n=0}^\infty \sum_{m=0}^\infty 
    \frac{G^n}{n!} \frac{E^m}{m!} {b^\dag}^n \frac{1}{F^{b^\dag b}} b^m \rho_S(0) {b^\dag}^m \frac{1}{F^{b^\dag b}} b^n.
\end{align}
Here, the time-dependent coefficients $E$, $F$, and $G$ are given by 
\begin{align}
\label{Eq: time-dep coeffs-1}
     & E = \frac{2}{F}(n_{th}+1) \sinh( \Gamma t/2)
     \\
    & F = \cosh( \Gamma t/2) + (2 n_{th}+1)\sinh( \Gamma t/2) 
    \\
    \label{Eq: time-dep coeffs-3}
     & G = \frac{2}{F}n_{th} \sinh( \Gamma t/2). 
\end{align}

To analyze the impact of decoherence and dissipation produced by TLSs, we compute the probability that the system is found in the initial phonon state $|\psi(0)\rangle$ at a later time. These dynamics are described by the fidelity 
$\mathcal{F}(t) =\langle \psi(0)| \rho_S(t) |\psi(0)\rangle$ \cite{nielsen2010quantum}. Using Eq. \eqref{Eq: rho_red}-\eqref{Eq: time-dep coeffs-3}, we find
\begin{align}
    \mathcal{F}(t) = \frac{e^{\Gamma t/2}}{F}\sum_{n=0}^\infty \sum_{m=0}^\infty 
    \frac{G^n}{n!} \frac{E^m}{m!} \left| \langle\psi(0)|{b^\dag}^n \frac{1}{F^{b^\dag b}} b^m |\psi(0)\rangle \right|^2.
\end{align}
which can be used to see how the stored quantum state degrades with time. At short times $\Gamma t\ll 1$, the fidelity approximately decays exponentially according to 
\begin{align}
    \mathcal{F}(t) \approx \exp\left\{ - \Gamma\left( (2 n_{th} +1) (\langle b^\dag b\rangle - |\langle b \rangle |^2) + n_{th}\right) t \right\} 
\end{align}
where $\langle ...\rangle$ denotes an expectation value with respect to the initial state. 
Using this expression, the storage time for a phononic quantum memory that is impacted by TLSs can be estimated. Assuming that the quantum information is lost once the fidelity falls below a threshold value $\mathcal{F}_{th}$, the storage time $T_{\mathcal{F}_{th}}$ can be estimated as
\begin{align}
\label{Eq: fidelity lifetime}
    T_{\mathcal{F}_{th}} \approx -\frac{\ln(\mathcal{F}_{th})}{\Gamma\left( (2 n_{th} +1)( \langle b^\dag b\rangle - |\langle b \rangle |^2) + n_{th}\right)} 
\end{align}
where the validity of this approximation requires $T_{\mathcal{F}_{th}} \Gamma \ll 1$.  

\subsection{Decoherence in quartz $\mu$BARs}

The operation and control of phonons in crystalline quartz resonators has now reached the quantum regime
\cite{doeleman2023brillouin,diamandi2025optomechanical}. While crystalline in nature, disorder at the resonator surfaces can host TLSs, and lead to excess loss and decoherence. As an example application of this model, we calculate the storage time for the initial phonon state $|\psi(0)\rangle = \alpha|0\rangle + \beta|1\rangle$, which is proposed for qubit encoding in quantum computing schemes utilizing hybrid architectures \cite{hann2019hardware}.

We begin this analysis by computing the coupling rates $g_j$ and $g_{0,j}$. For simplicity, we consider the fundamental dilatational mode of a confocal $\mu$BAR where the anisotopy of the crystal is neglected. In the limit where the crystal length $L$ is much less than the Rayleigh range and the beam waist is much greater than the phonon wavelength, diffraction can be neglected, and the phonon mode profile can be approximated by
\begin{align}
    {\bf u}_q \approx 
    -\sqrt{\frac{4}{\rho_0 \pi L w^2}} e^{-\frac{r^2}{w^2}}\cos(q z) \hat{z} 
\end{align}
where $w$ is the beam waist, $r$ is the radial coordinate parallel to the surfaces, $q = \pi m/L$ is the spatial frequency where $m$ is an integer, and the resonator is located between $z=0$ and $z=L$. For this mode, the coupling rates are given by 
\begin{align}
   \left. \begin{array}{cc}
         & g_j  \\
         & g_{0j}
    \end{array} \! \right\}= \sqrt{\frac{2}{\hbar \pi \rho_0 L w^2}}\frac{1}{E}
   \left\{ \! \begin{array}{cc}
          \Delta \\
         \Delta_0
    \end{array} \!
   \right\}
    e^{-\frac{r^2}{w^2}} \gamma_L q \sin q z.
\end{align}

Owing to the mesoscopic ($\sim$ mm) dimensions, the decay rates are given by the bulk system results but scaled by the effective fill fraction $f$ when the TLSs reside within a thin surface layer of thickness $\ell$
\begin{align}
\label{Eq: f effective}
    f = \frac{2}{L}\int_0^\ell dz \ \sin^2 qz = \frac{\ell}{L}\left(1-{\rm sinc}(2 q \ell) \right),
\end{align}
leading to expressions for $\Gamma_{res}$ and $\Gamma_{rel}$ given by
\begin{align}
 \Gamma_{rel} \approx f\frac{\pi^3}{24}\frac{P \gamma_L^2}{\rho_0^2 \hbar^4 v_L^2}\sum_\eta\frac{\gamma_\eta^2}{v_\eta^5}(k_B T)^3 
 \\
 \Gamma_{res} \approx f\frac{\pi P \gamma_L^2}{\rho_0 v_L^2} \Omega \tanh\left(\frac{\hbar \Omega}{2 k_B T} \right)
\end{align}
where $v_\eta$ ($\eta = L$ or $\eta = T$) is the speed of sound for $\eta$-polarized sound waves 
(see \cite{phillips1987two} \& \cite{behunin2016dimensional} for further details). 

Equation \eqref{Eq: f effective} shows that dissipation caused by TLSs can be reduced if the layer thickness $\ell$ is much less than the phonon wavelength, i.e., $q\ell \ll 1$. Because the strain vanishes at the surfaces ($z=0$ and $z=L$) for this example system, the TLS-induced dissipation is highly suppressed in this region according to the effective fill fraction
\begin{align}
\label{Eq: f effective app}
    f \approx \frac{2 \ell^3 q^2}{3 L}. 
\end{align}

\begin{figure}
    \centering  \includegraphics[width=\linewidth]{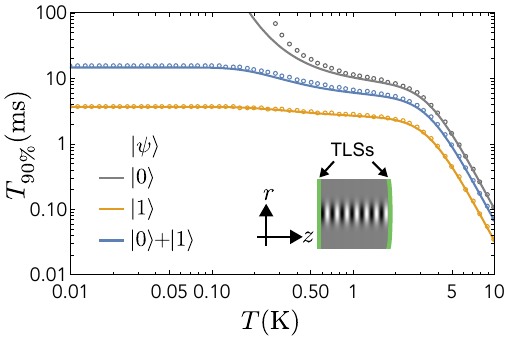}
    \caption{Lifetime of phonon fidelity vs. temperature for $|\psi(0)\rangle = |0\rangle$ (gray), $|\psi(0)\rangle = (|0\rangle+ |1\rangle)/\sqrt{2}$ (blue), and $|\psi(0)\rangle = |1\rangle$ (orange). Plots of $T_{90\%}$ are obtained using Eq. \eqref{Eq: fidelity lifetime} with $\mathcal{F}_{th} = 0.9$ (solid lines) and by numerically solving $\mathcal{F}(T_{90\%}) = 0.9$ (open circles). We assume the TLSs reside within a layer of thickness $\ell \approx 20$nm at the surface of the quartz resonator of length $L = 2.5$ mm (inset) with an effective fill fraction $f = 2 \times (3.78 \times 10^{-8})$ ($\times 2$ for both surfaces) (phonon wavevector $q = (2\pi) 2.14 \times 10^6$m$^{-1}$). For the TLSs in the surface layer, we use parameters  $P = 5.45 \times 10^{44}$ (J m$^3$)$^{-1}$, $\gamma_L = \sqrt{2}\gamma_T =$ 1 eV, $\rho_0 =$ 2620 kg/m$^3$, $v_L$ = 5900 m/s, and $v_T$ = 3500 m/s. The phonon frequency is assumed to be $12.6$ GHz.}
    \label{fig: coherence time}
\end{figure}

Figure \ref{fig: coherence time} displays the storage time for a collection of quantum states within a quartz $\mu$BAW. These results show the time required for the fidelity to decay to $90\%$,  i.e., $T_{90\%}$, as a function of temperature. Here, TLSs are assumed to reside in $\ell = 20$ nm layers on left and right surfaces of the $\mu$BAW (inset shown in green). We use known TLS parameters for silica glass, such as the density of states and deformation potential \cite{golding1973nonlinear}. 

Although the dissipation $\Gamma_{res}$ increases with lower temperature, we find that the quantum state lifetime is maximized at low temperatures. This is because the thermal occupation of the phonons ($n_{th}$), contained in the transition rates Eqs. \eqref{Eq: decay down} \& \eqref{Eq: decay up}, decreases in a manner that precisely compensates for increases in the mechanical decay, leading $T_{90\%}$ to be insensitive to temperature in the range $k_B T < \hbar \Omega$.   

\section{Conclusion}
In the paper, we have derived the master equation for a selected phonon mode interacting with an ensemble of TLSs. Using techniques from open quantum systems and standard approximations, this master equation reduces to a Lindblad form that is amenable to exact solution. The transition rates we find connect directly with known TLS resonant and relaxation absorption processes. While resonant absorption by TLSs saturates at high-temperatures, TLSs are still a strong source of decoherence owing to activation of relaxation processes. As an illustrative example, we analyze the quantum dynamics of a phonon mode in $\mu$BAR system. This example shows two important results: (1) that although the mechanical damping produced by $\Gamma_{res}$ reaches a local maximum at $T=0$, the phonon coherence time is maximized at low temperatures, and (2) the coupling to TLSs located near strain nodes is highly suppressed. These results provide the tools to calculate the quantum dynamics, and estimate the lifetime for exotic quantum states, of phonons in emerging systems.     

{\it Acknowledgments}---
This work was supported by NSF Awards No. 2145724 and No. 2427169.

\bibliography{refs}

 \appendix

 \section{Imaginary terms in the master equation}
\label{App: imaginary terms}

In Sec.~\ref{sec: master equation}, we introduced the susceptibility functions
\begin{align}
\chi_\pm(\Omega) \equiv \int_{0}^{\infty} dt \, \mathcal{G}(\pm t)\, e^{i\Omega t},
\end{align}
and obtained the master equation in the form
\begin{align}
       \dot{\rho}_S =
 2 {\rm Re}[\chi_+(\Omega)] b \rho_S b^\dag - 
 \chi_+(\Omega) b^\dag b\rho_S - \chi^*_+(\Omega)\rho_S b^\dag b \nonumber \\
 +2 {\rm Re}[\chi_+(-\Omega)] b^\dag \rho_S b - 
 \chi_+(-\Omega) b b^\dag \rho_S - \chi^*_+(-\Omega)\rho_S b b^\dag.
 \label{eq:ME-Lindblad1}
\end{align}
As seen in Sec. \ref{sec: master equation}, the real parts result in phonon dissipation, decoherence, and thermalization. Here, we isolate the imaginary parts and show that they generate dynamics equivalent to a frequency shift of the phonon mode. Writing
\begin{align}
\chi_+(\pm\Omega) = \chi_R(\pm\Omega) + i\,\chi_I(\pm\Omega),
\end{align}
where $\chi_R=\mathrm{Re}[\chi_+]$ and $\chi_I=\mathrm{Im}[\chi_+]$, and expanding each complex term in \eqref{eq:ME-Lindblad1} as such and neglecting real terms, one finds that the full imaginary contribution is
\begin{align}
\left.\dot{\rho}_S\right|_{\mathrm{imag}}
=
-i\Big(\chi_I(\Omega)+\chi_I(-\Omega)\Big)\,[b^\dag b,\rho_S].
\label{Eq:imag_total}
\end{align}
Equation~\eqref{Eq:imag_total} is of von Neumann form,
\begin{align}
\left.\dot{\rho}_S\right|_{\mathrm{imag}}
=
-\frac{i}{\hbar}[H_{\mathrm{shift}},\rho_S],
\end{align}
with the effective Hamiltonian
\begin{align}
H_{\mathrm{shift}}
=
\hbar\,\delta\Omega \; b^\dag b,
\qquad
\delta\Omega \equiv \chi_I(\Omega)+\chi_I(-\Omega).
\label{Eq:H_shift_def}
\end{align}
Thus, the imaginary parts of $\chi_+(\pm\Omega)$ do not produce decoherence; they renormalize the phonon frequency:
\begin{align}
\Omega \;\longrightarrow\; \Omega + \delta\Omega .
\end{align}
In the main body, we work in a rotating frame defined by
\begin{align}
H_0' = H_0 + \hbar\,\delta\Omega\, b^\dag b,
\end{align}
so that the term generated by $H_{\mathrm{shift}}$ can be canceled and the master equation only contains the dissipative (real-part) contributions. 

\section{Derivation of the Pauli operator correlation functions}
\label{App: TLS correlation functions}

The adjoint master equation for an operator $O$ is given by
\begin{align}
    \dot{O} = \frac{i}{\hbar}[H,O] + \sum_k \gamma_k (A_k^\dag O A_k -\frac{1}{2}\{A_k^\dag  A_k,O\})
\end{align}
where $A_k$ are the collapse operators and $\gamma_k$ are the transition rates \cite{breuer2002theory}. For a TLS coupled to a heat bath that can absorb, excite, and dephase, there are three relevant collapse operators
\begin{align}
    & A_1 = \sigma_- \quad \quad  \gamma_1 = \gamma_0 n_{th}
    \\
    & A_2 = \sigma_+ \quad \quad  \gamma_2 = \gamma_0 (n_{th}+1)
    \\
    & A_3 = \sigma_z \quad \quad \ \gamma_3 = \frac{1}{2 T_\phi}
    \end{align}
    where the excited state decay rate $T_1^{-1} = \gamma_0(2 n_{th}+1)$ and $T_\phi$ is the dephasing rate. The TLS decoherence rate $1/T_2$ is given by 
    \begin{align}
        \frac{1}{T_2} = \frac{1}{2 T_1}+\frac{1}{T_\phi}.
    \end{align}

Using the Pauli operator identities
\begin{align}
\label{Eq: Pauli-1}
    & \sigma_\mp\sigma_\pm = \frac{1}{2}(1\mp\sigma_z)
    \\
    \label{Eq: Pauli-2}
    &\sigma_\pm \sigma_z = \mp \sigma_\pm
    \\
    \label{Eq: Pauli-3}
    & \sigma_\pm \sigma_\pm = 0 \quad {\rm and} \quad \sigma_z^2 = 1
\end{align}
and the collapse operators and decay rates defined above, we obtain the adjoint master equations given below
\begin{align}
\label{Eq: adjoint master eq1}
    & \dot{\sigma}_z = - \frac{1}{T_1}(\sigma_z - w_0)
    \\
    \label{Eq: adjoint master eq2}
    & \dot{\sigma}_\pm = \left(\pm \frac{i}{\hbar}E - \frac{1}{T_2}\right)\sigma_\pm
\end{align}
where $w_0$ is the thermal equilibrium value for $\sigma_z$. 
By formally solving Eqs. \eqref{Eq: adjoint master eq1} \& \eqref{Eq: adjoint master eq2}, the two-time TLS correlation functions defined in Eq. \eqref{Eq: correlation function} can be evaluated. Using the solutions for $\sigma_z(t)$ and $\sigma_x(t)$ and assuming $t>0$, we find 
\begin{align}
    {\rm tr_R}\{\rho_R \sigma_{z}(t)\sigma_{z}(0) \}  = &{\rm tr_R}\{\rho_R (\sigma_{z}(0)e^{-t/T_1} 
    \\ \nonumber & \quad + w_0(1-e^{-t/T_1})) \sigma_{z}(0) \}   
    \\
    {\rm tr_R}\{\rho_R \sigma_{x}(t)\sigma_{x}(0) \}  = & {\rm tr_R}\{\rho_R (\sigma_{-}(0)e^{-(iE+1/T_2)t} 
    \\ & \quad + \sigma_{+}(0)e^{-(-iE+1/T_2)t}) \sigma_{x}(0) \}. \nonumber
\end{align}
These results reduce to Eqs. \eqref{Eq: two-time correlation functions} and \eqref{Eq: two-time correlation functions-2} when the equilbrium expectation values $ {\rm tr_R}\{\rho_R \sigma_z^2(0) \} = 1$, ${\rm tr_R}\{\rho_R \sigma_z(0) \} = w_0$, ${\rm tr_R}\{\rho_R \sigma_-(0) \sigma_x(0)\} = (1-w_0)/2$, and ${\rm tr_R}\{\rho_R \sigma_+(0) \sigma_x(0)\} = (1+w_0)/2$ are used, as well as Eqs. \eqref{Eq: Pauli-1}-\eqref{Eq: Pauli-3} and the definition $\sigma_x = \sigma_- + \sigma_+$.

\end{document}